\documentstyle[prl,aps]{revtex}
\begin{document}
\title{Quantum Hall Effect of Hard-Core Bosons}
\author{J.K. Jain$^1$ and Sumathi Rao$^{1,2}$}
\address{1. {\em Department of Physics, State University of New York
at Stony
Brook, Stony Brook, New York 11794-3800}\\
2. {\em Institute of Physics, Sachivalaya Marg, Bhubaneswar, 751005,
India}}
\date{\today}
\maketitle
\begin{abstract}

Motivated by a mean-field approach, which has been
employed for anyon superfluidity and the fractional quantum Hall
effect, the quantum Hall effect (QHE) of
hard-core bosons is investigated. It is shown that
QHE is possible {\em only} in the thermodynamic limit. The filling
factors where QHE may be expected are obtained with the help
of two adiabatic schemes.

\end{abstract}

\pacs{73.40.Hm}

The wave functions
of many non-interacting anyons
cannot be written in terms of single particle wave functions.
Fortunately, however, anyons can be represented as
fermions or bosons with gauge flux attached to them \cite{wilczek}.
For the definition of statistics, anyons must be hard-core,
i.e., their wave function must vanish whenever two anyons
coincide. As a result, the fermions or bosons must also
satisfy the hard-core property. This property is automatically obeyed
by (spinless) fermions, but has to be imposed upon bosons, and will
play a crucial role in the considerations below.
Moreover, since we start with non-interacting anyons,
the bosons and fermions are also non-interacting (except for the
statistical gauge interaction).
It will be assumed throughout this paper, unless otherwise specified,
that bosons are non-interacting, except for the hard-core constraint.
The amount of  flux carried by the fermions
or bosons is chosen so that the particles acquire  the correct phases as
they wind around each other. The real usefulness of this representation
becomes evident in a clever mean-field approach \cite{laughlin},
in which the gauge flux is adiabatically delocalized,
so that, at a mean-field level, the problem
reduces to that of particles of regular statistics
in a uniform magnetic field. There is a
large body of literature suggesting that incompressibility in the
mean-field fermion or boson system implies superfluidity
in the anyon system \cite{fhl,wen}. Incompressible states of non-interacting
fermions have also been related to the fractional quantum Hall effect
(FQHE) of interacting fermions by a similar mean-field approach \cite{jain89}.
Therefore, incompressible states of non-interacting
fermions and bosons in a magnetic field are of special interest.

The problem of non-interacting fermions in the presence of an
external magnetic field is exactly solvable.  In particular, at
filling factors given by  $\nu\equiv \phi_{0} \rho /B =n$,
where $\phi_{0}=hc/e$ is the fundamental flux quantum, $\rho$ is the
fermion density, $B$ is the external magnetic field, and $n$ is an
integer, the ground state is unique and incompressible.
It is simply the state with $n$ filled Landau levels (LLs),
which will be denoted by
$\Phi_{n}$. These states relate to anyons of statistics
$\theta=1+n^{-1}$
(which are equivalent to $\theta=1-n^{-1}$ due to the periodicity of
$\theta$),
where the statistics $\theta$ is defined so that an exchange of two
anyons produces a phase factor of $e^{i\pi \theta}$.

For a study of anyon superfluidity, the
mapping to bosons may seem a more natural starting point. There
have been relatively few attempts to attack the problem from the boson
end \cite{wen}, mainly because our understanding of the QHE of
non-interacting hard-core bosons is not satisfactory.
It was argued in Ref. \cite{wen}, with the help of a largely intuitive
semi-classical picture, that the boson system is also
incompressible at {\em integer} filling factors, which would result
in anyon superfluidity for $\theta=1/n$.
However, the hard-core property was not treated properly in these
arguments, nor was the boson statistics. In contrast, another
study \cite{zhang} claims that
hard-core bosons do not exhibit {\em any} QHE whatsoever.
If true, this would make the boson mean-field approach totally
irrelevant for the issue of anyon superfluidity.

The objective of this paper is two-fold. First, we show that the
argument of Ref. \cite{zhang} is not applicable {\em in the thermodynamic
limit}, and, as a matter of principle,
QHE of hard-core bosons is indeed possible.
We then use two adiabatic schemes to determine the filling
factors where boson QHE may be expected.
It should be emphasized that the problem of the QHE
of {\em non-interacting} (hard-core) bosons
is different from the problem of the
{\em fractional QHE} (FQHE) of {\em interacting} bosons, which is
relatively well understood \cite{haldane,xie}. In the former case, QHE
{\em must} occur due to cusps in the {\em kinetic energy} as a
function of filling factor, analogous to the {\em integer} QHE
(IQHE) of fermions. We will see that there are many filling factors
where QHE occurs for interacting bosons, but not for non-interacting
bosons, {\em and vice versa}.

Let us first state the conventions and some well-known
results that will be used in the rest of the paper.
All lengths will be expressed in units of the magnetic length,
$l=\sqrt{\hbar c/eB}$. The energies will be expressed in units of the
cyclotron energy, $\hbar\omega_{c}=\hbar eB/m_{e}c$, $m_{e}$ being the
particle mass, and the zero of the energy will be set at the lowest LL
(LLL). Thus, the energy
of a state in the $n$th LL is $n$, $n=0$ being the LLL.  $\Phi$ will
be used for  an antisymmetric fermion wave function and
$\Psi$ for a symmetric boson wave function. The subscript
of $\Phi$ or $\Psi$ will denote the filling factor.
The (unnormalized) single particle eigenstates in the
LLL are given by $z^m\exp[-|z|^2/4]$
where $z=x-iy$ denotes the position of the particle \cite{laughlin83}.
Thus, apart from the exponential factor,
$\exp[-\sum_{j}|z_{j}|^2/4]$, which will be often suppressed
for simplicity, the wave function of any many-particle
state strictly in the LLL must  be an analytic function of
$z_{j}$. The largest power of
$z_{j}^*$ allowed in a state restricted within the lowest $p$ LLs is
$p-1$. Thus, it is possible to tell how many LLs are involved
in a wave function simply by looking at the factor multiplying
the exponential.

Our first task is to show that QHE can actually occur for
hard-core bosons. As will become clear later,
in order for QHE to occur,
a crucial role is played by the assumption that bosons occupy as few LLs
as possible to minimize their (kinetic) energy. While this sounds
reasonable, it was argued in Ref. \cite{zhang} that, by making use of
{\em all} LLs, bosons can lower their energy much more -- in
fact, to such an extent that there is no QHE!
The argument can be summarized as follows. Construct
a boson wave function in the LLL without worrying about the
hard-core constraint. This, of course,
can be done at arbitrary filling factor.
Now, multiply it by $|D|^{2\eta}$ to obtain
\begin{equation}
\Psi_{\eta}=|D|^{2\eta} F_{s}[\{z_{i}\}]
\exp[-\sum_{j}|z_{j}|^2/4]\;\;,
\end{equation}
\begin{equation}
D\equiv \prod_{j<k}(z_{j}-z_{k})\;\;,
\end{equation}
where $F_{s}[\{z_{i}\}]$ is a symmetric function of $z_{j}$.
$\Psi_{\eta}$ clearly satisfies the hard-core constraint.
It was shown in Ref.
\cite{zhang} that, for small $\eta$, the kinetic energy {\em per
particle} of this state,
$E_{k}^{\eta}$, is proportional to $\eta$.  {\em For a finite
system}, this indeed proves that, by choosing a
sufficiently small $\eta$, it is possible to construct a state at
arbitrary $\nu$ with energy as close to zero as one pleases, and, as
a result, there is no QHE.

In the thermodynamic limit, however, the situation is more subtle. It
is now crucial to know how $E_{k}^{\eta}$ behaves as a function of
$N$, the number of bosons. If it turned out that $E_{k}^{\eta}\sim \eta $,
the state $\Psi_{\eta}$ would have arbitrarily low energy
and the absence of QHE would continue to hold in the limit
$N\rightarrow\infty$.  However, we now show that
\begin{equation}
E_{k}^{\eta}=O(\eta N)\;\;.
\label{ke}
\end{equation}
In terms of $z$ and $z^*$, the kinetic energy operator is given by
$$H=\frac{1}{2}\sum_{i=1}^{N}(-4\partial_{i} \partial_{i}^*+
\frac{1}{4} z_{i}z_{i}^*-z_{i}\partial_{i}+z_{i}^*\partial_{i}^*-1)
\;\;,
$$ where $\partial_{i}\equiv \partial/\partial z_{i}$.
Let us first specialize to the case when $F_{s}[\{z_{i}\}]=1$.
Application of $H$ on $\Psi_{\eta}$ then gives
$$H\Psi_{\eta}=e^{-\frac{1}{4}\sum_{j}|z_{j}|^2}\;\sum_{i}
(-2\partial_{i}+z_{i}^*)\partial_{i}^*
D^{\eta}D^{*\eta}=H'\Psi_{\eta}\;,
$$ where
$$H'=\sum_{i}[-2\eta^2 \sum_{l,m} \frac{1}{(z_{i}-z_{l}) (z_{i}^*-
z_{m}^*)} + \eta \sum_{l} \frac{z_{i}^*}{z_{i}^*-z_{l}^*}]\;\;,
$$
with $l,m\neq i$. The kinetic energy is given by
the expectation value of $H'$.
The second term in $H'$ can be evaluated explicitly:
$$\eta \sum_{i\neq l} \frac{z_{i}^*}{z_{i}^*-z_{l}^*} =
\frac{\eta}{2} \sum_{i\neq l} [\frac{z_{i}^*}{z_{i}^*-z_{l}^*}+
\frac{z_{l}^*}{z_{l}^*-z_{i}^*}]=\frac{\eta}{2}N(N-1)\;\;.
$$
As this term cannot be canceled by the expectation value of the
first term in $H'$ (which is proportional to $\eta^2$),
Eq.\ (\ref{ke}) follows. (Note that the kinetic energy is guaranteed
to be positive.) In the general case, when $F_{s}\neq
1$, $H'$ contains a third term proportional to $\eta$.
But unless the expectation value of this term  cancels {\em exactly} the
second term, which we consider extremely unlikely, Eq.\ (\ref{ke})
remains true.  Thus, for arbitrarily small but {\em fixed} $\eta$,
the kinetic energy per particle of $\Psi_{\eta}$ is divergent in
the thermodynamic limit, $N\rightarrow \infty$, whereas  the
kinetic energy per particle of a state
restricted to a {\em finite} number of
lowest LLs is finite [O(1)].
Therefore, in the thermodynamic limit,
the state $\Psi_{\eta}$ is irrelevant, and the
hard-core constraint cannot be accomodated at an infinitesimal cost
in energy. Our usual intuition that low-energy states are obtained by
putting particles in as few Landau levels as possible
is then likely to be right
and QHE should be possible in principle.

We emphasize that the above conclusion requires
taking the thermodynamic limit while keeping $\eta$
fixed. The parameter $\eta$ is related to the
`size' of the bosons, which is roughly given by $e^{-1/\eta}$.
This can be seen by considering two particles at a distance $r$;
the wave function behaves as $r^{\eta}=e^{\eta \ln r}$, and differs
appreciably from unity when $r \leq e^{-1/\eta}$.
We will assume in the rest of the paper that it is valid to keep $\eta$
finite (though arbitrarily small).

There is no limit to the number of {\em ideal} bosons that can be
put in the LLL.
However, with the hard-core constraint, this no longer remains true.
The largest possible filling factor in the LLL is $\nu=1/2$, as
shown by Haldane \cite{haldane}. We repeat his argument here for
completeness.  In order to satisfy the hard-core requirement,
the boson wave function in the LLL, which is an analytic function of
$z_{j}$, must vanish whenever two coordinates are
identified. This requires it to be of the form
$D\Phi[\{z_{j}\}]$.
Since $\Phi$ is an analytic antisymmetric function, it
must also contain another factor of $D$.
Hence the wave function of hard-core bosons in the LLL has the form
$\Psi=D^2\;\Psi'$,
where $\Psi'$ is a symmetric analytic function.
The largest filling factor is $\nu=1/2$, obtained with
$\Psi'=1$, when the bosons form a Laughlin state \cite{laughlin83}.
In order to add more bosons, higher LLs must be
involved, which costs non-zero energy.
Thus, for hard-core bosons, the ground state energy is zero
for $\nu \leq 1/2$ and non-zero otherwise. This strongly suggests
that there is a cusp in the ground state energy at $\nu = 1/2$,
resulting in incompressibility and QHE at $\nu=1/2$.

In order to obtain other filling factors where QHE may occur, we
employ two adiabatic schemes. In each case, we start with a model
where the physics of QHE is more transparent, and then argue that
this model may be connected to the model of interest, namely that of
non-interacting bosons in an external magnetic field.

In the first scheme, we relax the hard-core
condition, and denote the energy cost of putting two bosons on top of
each other by $U$ (measured in units of the cyclotron energy).
The hard-core condition is obtained for
$U = \infty$. Now let us consider the situation when $U<<1$,
so that the bosons reside within the LLL. These do not in general
satisfy the hard-core condition, and their states will be denoted
by the superscript LLL. In this case, the bosons exhibit
{\em FQHE} \cite{haldane,xie}. In analogy to the fermion FQHE \cite
{jain89}, the hard-core bosons are likely to exhibit FQHE at
$\nu_{n}=n/(n\pm 1)$ and we claim that the exact
incompressible states are well represented by
\begin{equation}
\Psi_{n/(n\pm 1)}^{LLL}= {\cal P}\Psi_{n/(n\pm 1)} \;\;,
\end{equation}
\begin{equation}
\Psi_{n/(n\pm 1)} = D \Phi_{\pm n}\;\;,
\label{dnw}
\end{equation}
where ${\cal P}$ is the LLL projection operator.
These states are justified from the following two facts: (i)
Ref. \cite {dev} demonstrated that the LLL fermion FQHE states
at $\nu=n/(2n\pm 1)$ are well approximated by
$$\Phi_{n/(2n\pm 1)}^{LLL}=D {\cal P} \Psi_{n/(n\pm 1)}\;\;. $$
(ii) Xie {\em et al.} \cite {xie} showed that the exact fermion
state at $\nu=n/(2n\pm 1)$ and the boson state at $\nu=n/(n\pm 1)$
are related through the Jastrow factor, $D$.
It then follows that the exact boson state is
accurately given by ${\cal P}\Psi_{n/(n\pm 1)}$ \cite {coulomb}.

Yet another piece of information is essential for the adiabatic
scheme. This is that, despite some LL mixing,
$\Psi_{n/(n+1)}$ are predominantly
in the LLL (with the amplitude in higher LLs increasing with $n$).
This is expected because a) $\Phi_{n}$ involves only a finite number of
LLs, and b) the higher LL states in $\Phi_{n}$ are
multiplied by large powers of $z_{j}$ present in $D$ \cite {jain89,jain90}.
Detailed Monte Carlo calculations on the analogous fermion wave functions
have confirmed this expectation \cite {trivedi}.
The states at $\nu=n/(n-1)$, $\Psi_{n/(n-1)}$,
on the other hand, are not expected to
satisfy this property, since $z_{j}^*$ in $\Phi_{-n}=\Phi_{n}^*$ occurs
with arbitrarily large powers, implying that an infinite number of LLs
is involved.

Start now with the the true LLL FQHE state at
\begin{equation}
\nu=n/(n+1)\;\;,
\end{equation}
and  increase $U$. This results in
mixing with higher LLs. However, since it {\em is} possible for bosons
to avoid0each other completely with the help of only a slight amount
of LL mixing, as demonstrated by the wave function $\Psi_{n/(n+1)}$,
it is plausible that, at least for small $n$,
the $U<<1$ state will {\em adiabatically} evolve into the
$U=\infty$ state. In this case, the gap of the $U<<1$ state will
continuously evolve into a gap at $U=\infty$, leading to
incompressibility for hard-core bosons. While at $U<<1$
the gap is determined by the interaction strength $U$, at $U=\infty$
it must be proportional to $\hbar\omega_{c}$, the only energy scale
in the problem. Since we started from ${\cal P} \Psi_{n/(n+ 1)}$,
which is very close to a hard-core state $\Psi_{n/(n+ 1)}$, it is likely that
the $U=\infty$ state will be well approximated by $\Psi_{n/(n+ 1)}$.
The fate of the LLL FQHE states  ${\cal P} \Psi_{n/(n-1)}$ is not
as clear, but the most likely scenario is that
the gap will be destroyed at a finite $U$.
Note that {\em for the Coulomb interaction}, FQHE may also occur at
$\nu=n/[n(2m+1)\pm 1]$, with the incompressible state given by
$D^{2m}\Psi_{n/(n\pm 1)}$. This state is not relevant to the present
problem, since the wave function vanishes faster than required by the
hard-core condition, and is consequently degenerate with a
large number of other wave functions satisfying the hard-core property.

Thus, QHE is made plausible at filling factors
$\nu=n/(n+1)$, which are related to
anyons of statistics $\theta=1+ n^{-1}$. These values of $\theta$
are are identical to
those obtained from the fermion mean-field theory.  The states
$\Psi_{n/(n+1)}$ are also in a complete one-to-one
correspondence with the incompressible fermion states $\Phi_{n}$.

In the second adiabatic scheme, we construct an artificial  model
in which the energy of the $n$th LL is taken to be
\begin{equation}
E_{n}=n\;, \;\;\;\;\; n \geq K\;\;,
\end{equation}
\begin{equation}
E_{n}=\alpha n \;, \;\;\;\;\; n=0,1,...,K-1.
\end{equation}
Thus, the energies of the lowest $K$ LLs are treated as variable.
The `physical' problem corresponds to $\alpha=1$.

Let us first consider $K=2$, $\alpha=0$, when the lowest two LLs
are degenerate with zero energy, and ask: (a) What is the largest
filling factor
where a zero energy state exists? (b) What is this state?
We start by constructing the most general hard-core wave function
strictly confined to the lowest two LLs. As before, it must vanish
as $r_{jk}^2$, when the distance between two bosons,
$r_{jk}\rightarrow 0$.
Due to the confinement within the lowest two LLs, at most one power
of $z_{j}^*$ or $z_{k}^*$ is allowed,
so one power of $r_{jk}^2$ must come from $(z_{j}-z_{k})$.
The boson wave function must therefore have the form
$\Psi_{\nu}= D\Phi_{\nu^*}$,
where $\Phi_{\nu^*}$ is a completely antisymmetric wave function
confined to the lowest two LLs. The filling factor of the product is
$\nu=\nu^*/({\nu^*}+1)$
and the largest possible value, $\nu=2/3$, is achieved when $\nu^*$
assumes its largest value, $\nu^*=2$. The boson ground state here
is given by $\Psi_{2/3}=D\Phi_{2}$, which is
the unique state at $\nu=2/3$ within the lowest two LLs
satisfying the hard-core property.
The boson system has zero energy for $\nu\leq 2/3$ and non-zero
energy for $\nu>2/3$. This suggests that  the $K=2$, $\alpha=0$
model exhibits QHE at $\nu=2/3$.

Now increase $\alpha$. Clearly, due to the
presence of a gap at $\alpha=0$, the system is likely to be insensitive
to small changes of $\alpha$. We believe that the gap will survive
all the way to the physical value $\alpha=1$. The reason is that the
main effect of increasing $\alpha$ is to reduce the higher LL
occupation of bosons in the actual ground state, but, as mentioned
before, $\Psi_{2/3}$ is already predominantly in the LLL.
Therefore, it is plausible that no significant rearrangement
of bosons will take place as $\alpha$ is increased from 0 to 1, and
the ground state at $\alpha=1$ is adiabatically connected to
$\Psi_{2/3}$ \cite{rezayi}.

Next we consider general $K$ and
look for the maximum filling factor where a hard-core state can be
constructed within the lowest $K$ LLs, which would have zero energy.
Let us assume that this  state is a product of two fermion states,
which would satisfy the hard-core as well as the symmetry requirements.
Then, for  $K=2m-1$, $m$ being an integer, the largest possible
filling factor where a zero energy state occurs is
\begin{equation}
\nu=\frac{m}{2}\;\;,
\label{ff1}
\end{equation}
where the zero energy ground state is given by
$\Phi_{m}^2$,
and for $K=2m$, it is
\begin{equation}
\nu=\frac{m(m+1)}{2m+1}\;\;,
\label{ff2}
\end{equation}
where the ground state is $\Phi_{m+1}\Phi_{m}$.
These states are of the type considered in Ref. \cite{jain89b}.
The occupation of higher ($n\neq 0$) LLs in these states
is expected to increase with increasing $m$, and consequently,
their adiabatic connection to the true $\alpha=1$ ground state
is more questionable for larger $m$.
This adiabatic scheme gives many new filling factors, and possibly
new anyon superfluid states, which do not have any counterparts
in either the fermionic mean-field theory or the first adiabatic
scheme.  Note that while the boson states in Eq.\ (\ref{dnw})
can be interpreted as the IQHE states of ``composite bosons", where
a composite boson is a fermion carrying one flux quantum, the
other states do not lend themselves to a similar
interpretation. Also, while the the excitations of most of
the states obtained above are fractionally charged and have fractional
statistics \cite {halperin}, the states $\Phi_{m}^2$ possess
quasiparticles of {\em non-abelian} statistics \cite {wen2}.

The filling factors $1/2$ and $2/3$ are obtained in both
adiabatic schemes, and correspond to identical microscopic states in
both cases, since $\Phi_{1}=D$. For this
reason, we have more confidence in QHE at these two filling factors.
Numerical investigation of these as well as
the simplest next states in the two schemes, $D\Phi_{3}$ at $\nu=3/4$
and $\Phi_{2}^2$ at $\nu=1$, is desirable. Note that
boson QHE at integer filling factors is deemed possible,
but the structure of the incompressible states here is rather subtle,
and distinct from that envisaged in the works of Ref.\cite{wen}.

In conclusion, we have identified the importance of the thermodynamic
limit for the issue of the QHE of non-interacting
hard-core bosons, and used two adiabatic schemes to
determine the filling factors where QHE may be expected.
An essential role is played by the hard-core nature of bosons, which
causes a mixing with higher LLs for $\nu>1/2$; QHE is expected
whenever it becomes necessary to occupy a new LL. While
further work is needed for a confirmation of our
approach, it is remarkable that concepts as exotic as
fractional, and even non-abelian,
statistics are possibly relevant for a system as simple as that
of hard-core bosons in a uniform magnetic field.

This work was supported in part by the National Science
Foundation under Grant No. DMR90-20637. One of us (S.R.) thanks the
theory groups at Fermilab and SLAC for hospitality during the
completion of this work.


\begin{thebibliography}{99}
\bibitem{coulomb} These studies were done for the Coulomb
interaction. However, due to the insensitivity of the  incompressible
states to the actual form of the interaction, the results
are expected to hold for short-range contact interaction as well.
\end{thebibliography}
\end{document}